\documentclass[aps,prl,twocolumn,showpacs,showkeys,floatfix,footinbib,longbibliography,superscriptaddress]{revtex4-1}

\usepackage{graphics,amssymb,amsmath,epsfig,color}
\usepackage{amsmath}
\usepackage{amssymb}
\usepackage{graphicx}
\usepackage{epstopdf}
\usepackage{bm}
\usepackage{color}
\usepackage{relsize}
\usepackage{braket}
\usepackage[caption=false]{subfig}
\usepackage{lipsum}
\usepackage{hyperref}
\usepackage{fixltx2e}
\usepackage{amsfonts}
\usepackage{bbm}
\usepackage[all]{hypcap}
\usepackage{xcolor}
\usepackage{braket}

\usepackage{amssymb}
\usepackage{mathrsfs}
\usepackage{amsthm}

\begin{document}
\title{Signatures of enhanced spin-triplet superconductivity induced by interfacial properties}
\author{Chenghao Shen}
\affiliation{Department of Physics, University at Buffalo, State University of New York, Buffalo, NY 14260, USA}
\author{Jong E. Han}
\affiliation{Department of Physics, University at Buffalo, State University of New York, Buffalo, NY 14260, USA}
\author{Thomas Vezin}
\affiliation{Department of Physics, University at Buffalo, State University of New York, Buffalo, NY 14260, USA}
\affiliation{Laboratoire des Solides Irradies, Ecole Polytechnique, Universite Paris-Saclay, F-91767 Palaiseau Cedex, France}
\author{Mohammad Alidoust}
\affiliation{Department of Physics, Norwegian University of Science and Technology, N-7491 Trondheim, Norway}
\author{Igor \v{Z}uti\'c}
\affiliation{Department of Physics, University at Buffalo, State University of New York, Buffalo, NY 14260, USA}

\begin{abstract}
While spin-triplet pairing remains elusive in nature, there is a growing effort to realize proximity-induced equal-spin triplet superconductivity in junctions with magnetic regions or an applied magnetic field and common $s$-wave superconductors. To enhance such spin-triplet contribution, it is expected that junctions with a weak interfacial barrier and strong spin-orbit coupling are desirable. Intuitively, a weak interfacial barrier enables a robust proximity-induced superconductivity and strong spin-orbit coupling promotes spin mixing, converting spin-singlet into spin-triplet superconductivity. In contrast, we reveal a nonmonotonic spin-triplet contribution with the strength of the interfacial barrier and spin-orbit coupling. This picture is established by considering different signatures in conductance and superconducting correlations, as well as by performing self-consistent calculations. As a result, we identify a strongly enhanced spin-triplet superconductivity, realized for an intermediate strength of interfacial barrier and spin-orbit coupling. In junctions with magnetic regions, an enhanced spin-triplet superconductivity leads to a large magnetoanisotropy of conductance and superconducting correlations. This picture of an enhanced spin-triplet superconductivity is consistent with experiments demonstrating a huge increase in the conductance magnetoanisotropy, which we predict can be further enhanced at a finite bias.
\end{abstract}
\maketitle

\section{I. Introduction}
\label{sec:Intro}
There is a continued quest to identify systems which would support spin-triplet superconductivity~\cite{Amundsen2024:RMP}.
On one hand it could allow for coexistence of ferromagnetism and superconductivity with long-range proximity effects where such superconductivity extends into the ferromagnet over microscopic lengths, just as for the common spin-singlet superconductivity in a normal metal~\cite{Eschrig2015:RPP,Valls:2022,Bergeret2005:RMP}. The underlying equal-spin Cooper pairs would 
provide fascinating opportunities for superconducting spintronics~\cite{Eschrig2011:PT,Linder2015:NP,Eschrig2015:RPP,Han2020:NM,Cai2023:AQT}, since they carry spin current and angular 
momentum. On the other hand, spin-triplet superconductivity is sought for topological superconductivity and implementing fault-tolerant quantum computing through control of the resulting Majorana bounds states~\cite{Kitaev2001:PU,Laubscher2021:JAP,Gungordu2022:JAP,Flensberg2021:NRM}.

Despite decades of efforts to identify spin-triplet superconductivity, materials candidates and its characteristic signatures remain debated. This is exemplified 
in Sr$_2$RuO$_4$~\cite{Maeno1994:N}, long held to be the prime candidate for a $p$-wave spin-triplet superconductivity~\cite{Mackenzie2003:RMP} and proposed as a platform for Majorana states~\cite{Sarma2006:PRB}. In contrast, the claimed definitive phase-sensitive spin-triplet signatures~\cite{Nelson2004:S,Rice2004:S} were suggested could also come from spin-singlet superconductivity~\cite{Zutic2005:PRL}, while recent experiments argue against spin-triplet superconductivity in Sr$_2$RuO$_4$, even involving its discoverer, Y. Maeno~\cite{Sharma2020:PNAS,Petsch2020:PRL}.

Unlike seeking elusive spin-triplet superconductivity in a single material, such as Sr$_2$RuO$_4$, there is a growing effort to realize it through proximity effects. This can be seen both in superconducting spintronics as well as in the search for topological superconductivity and Majorana states~\cite{Eschrig2011:PT,Linder2015:NP,Martinez2016:PRL,
Eschrig:2019,Fu2008:PRL,Keizer2006:N,Banerjee2014:NC,Jeon2018:NM,Johnsen2019:PRB,Lutchyn2010:PRL,Oreg2010:PRL,
Fornieri2019:N,Ren2019:N,Dartiailh2021:PRL,Setiawan2019:PRB,Alidoust2021:PRB}. Nevertheless, even in simple proximitized materials, transformed by proximity effects~\cite{Zutic2019:MT}, there are challenges in identifying the signatures of spin-triplet superconductivity. A prediction that a spin-triplet superconductivity 
and Majorana states are associated with quantized zero-bias conductance peak~\cite{Sengupta2001:PRB} has been used in many experiments~\cite{Mourik2012:S,Deng2012:NL,Das2012:NP,Deng2016:S}, but also shown to arise from spurious effects~\cite{Yu2021:NP} and 
invalidate experimental reports of Majorana states~\cite{Zhang2021:N}. 

Motivated by these developments we examine signatures of spin-triplet superconductivity from conductance and pair correlations that arise in simple ferromagnet/s-wave superconductor (F/S) junctions in the presence of interfacial spin-orbit coupling (SOC)~\cite{Hogl2015:PRL,Martinez2020:PRA,Cai2021:NC}. 
Such SOC and the resulting emergent interfacial spin-orbit fields are directly realized in junctions through structural inversion asymmetry~\cite{Zutic2004:RMP,Fabian2007:APS}, while the tunneling anisotropic magnetoresistance (TAMR) allows their experimental detection~\cite{Moser2007:PRL,Fabian2007:APS,Zutic2019:MT}. 
However, in the normal state, the observed 
MR is rather small (typically $<1\%$), even at low temperatures for magnetic junctions with large spin polarization~\cite{Gould2004:PRL}.

In comparing different signatures of spin-triplet superconductivity it is not obvious how they are related to each other and if they display similar trends with interfacial properties of F/S junctions.
A cautionary remark for this work is provided by a seemingly obvious concept of the spin polarization~\cite{Mazin1999:PRL}. 
However, a closer look reveals a number of subtleties since the measured spin polarization depends on the employed experimental probe~\cite{Mazin1999:PRL,Zutic2004:RMP}. In a ferromagnetic nickel a spin polarization from the photoemission can even have a different sign than that obtained from the conductance measurements~\cite{Mazin1999:PRL}.

While we have previously suggested that using zero-bias information about the equal-spin Andreev reflection can identify spin-triplet superconductivity~\cite{Vezin2020:PRB}, it is unclear if that simple signature is consistent with considering superconducting correlations. Such correlations provide a very different signature of spin-triplet superconductivity. They are not limited to zero-bias behavior and contain spatially-resolved information, used to identify long-range equal-spin triplet superconducting proximity effects, with and without SOC~\cite{Amundsen2024:RMP,Eschrig2011:PT,Linder2015:NP,Eschrig2015:RPP,Konschelle2016:PRB,Silaev2020:PRB,Bergeret2014:PRB}.

\begin{figure}[t]
	\centering
	\includegraphics*[trim=0.75cm 0.8cm 0.4cm 0.0cm,clip,width=8.4cm]{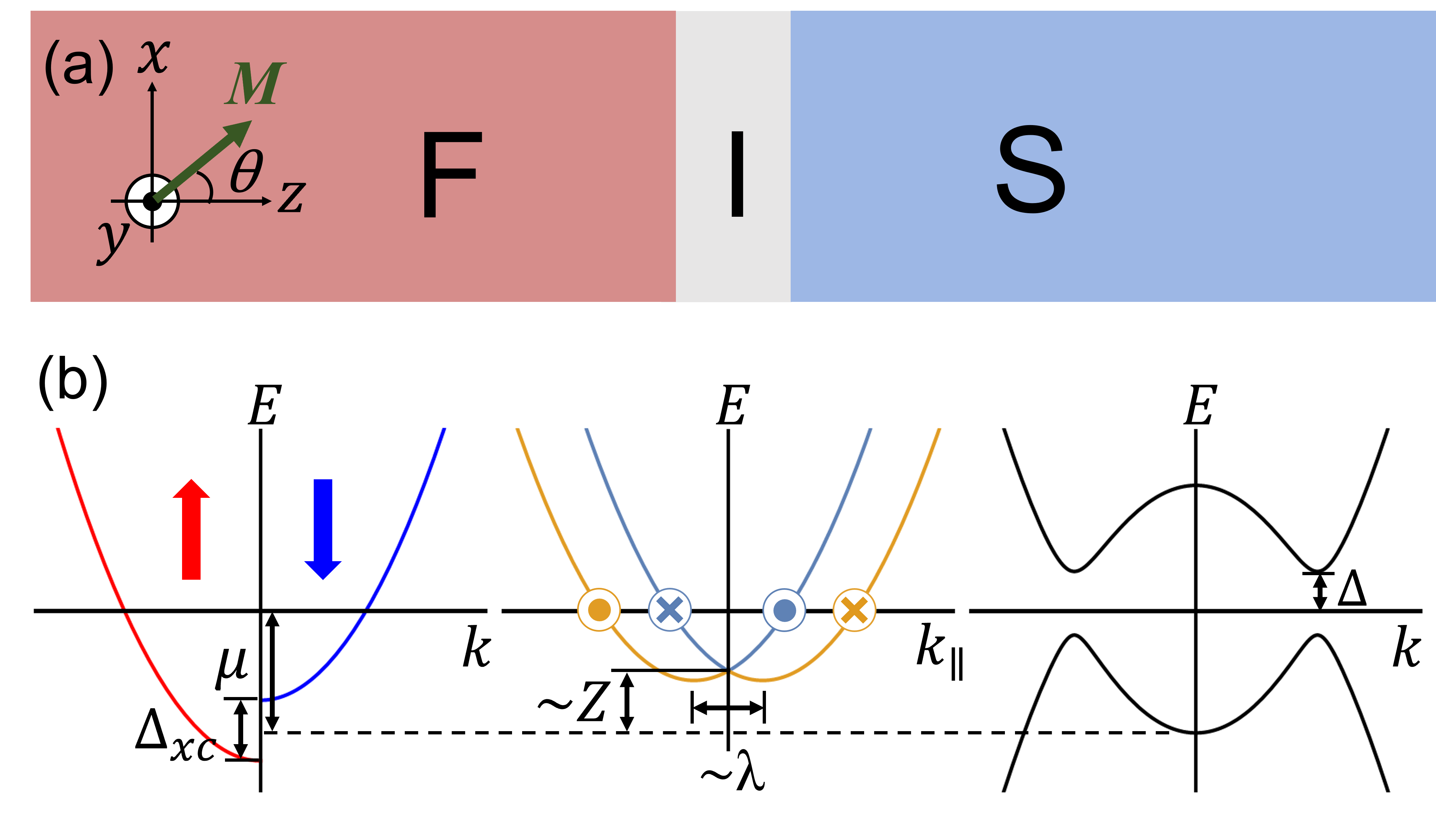}
	\caption{(a) Ferromagnet/superconductor (F/S) junction separated by an insulator (I) with potential and Rashba spin-orbit scattering (SOC). $\bm{M}$ is the magnetization and the current flows along $\bm{z}$. (b) The corresponding band structure is given for each region. Spins are denoted by arrows: In the F region red (blue) for parallel (antiparallel) to $\bm{M}$; with interfacial SOC, spins (dots, crosses) are parallel to the interface and perpendicular to the in-plane wave vector, $k_\parallel$. $\Delta_{xc}$ is the exchange splitting and $\mu$ the chemical potential. The barrier and SOC strengths are $Z$ and $\lambda$. In the S region the superconducting bands open a superconducting gap $\Delta$.}
	\label{fig:scheme}
\end{figure}

We consider the F/S geometry depicted in Fig.~\ref{fig:scheme}, where the dimensionless strength of the interfacial barrier and the Rashba SOC are parametrized by $Z$~\cite{Blonder1982:PRB,Zutic1999:PRBa} and $\lambda$~\cite{Hogl2015:PRL,Vezin2020:PRB}.
In contrast to common expectation that a strong SOC should be desirable for spin-triplet superconductivity, we reveal a more complex picture in which the SOC strength that maximizes the spin-triplet contribution nonmonotonically depends on the interfacial barrier in F/S junctions. 
Conversely, a weak interfacial barrier that enables a robust proximity-induced superconductivity seems suitable to enhance the spin-triplet superconductivity. 
Instead, we find that the most enhanced spin-triplet contribution is obtained for the interfacial barrier that nonmonotonically depends the SOC strength. 

\begin{figure}[t]
	\centering
	\includegraphics*[trim=0.0cm 0.4cm 1.2cm 0.6cm,clip,width=8.4cm]{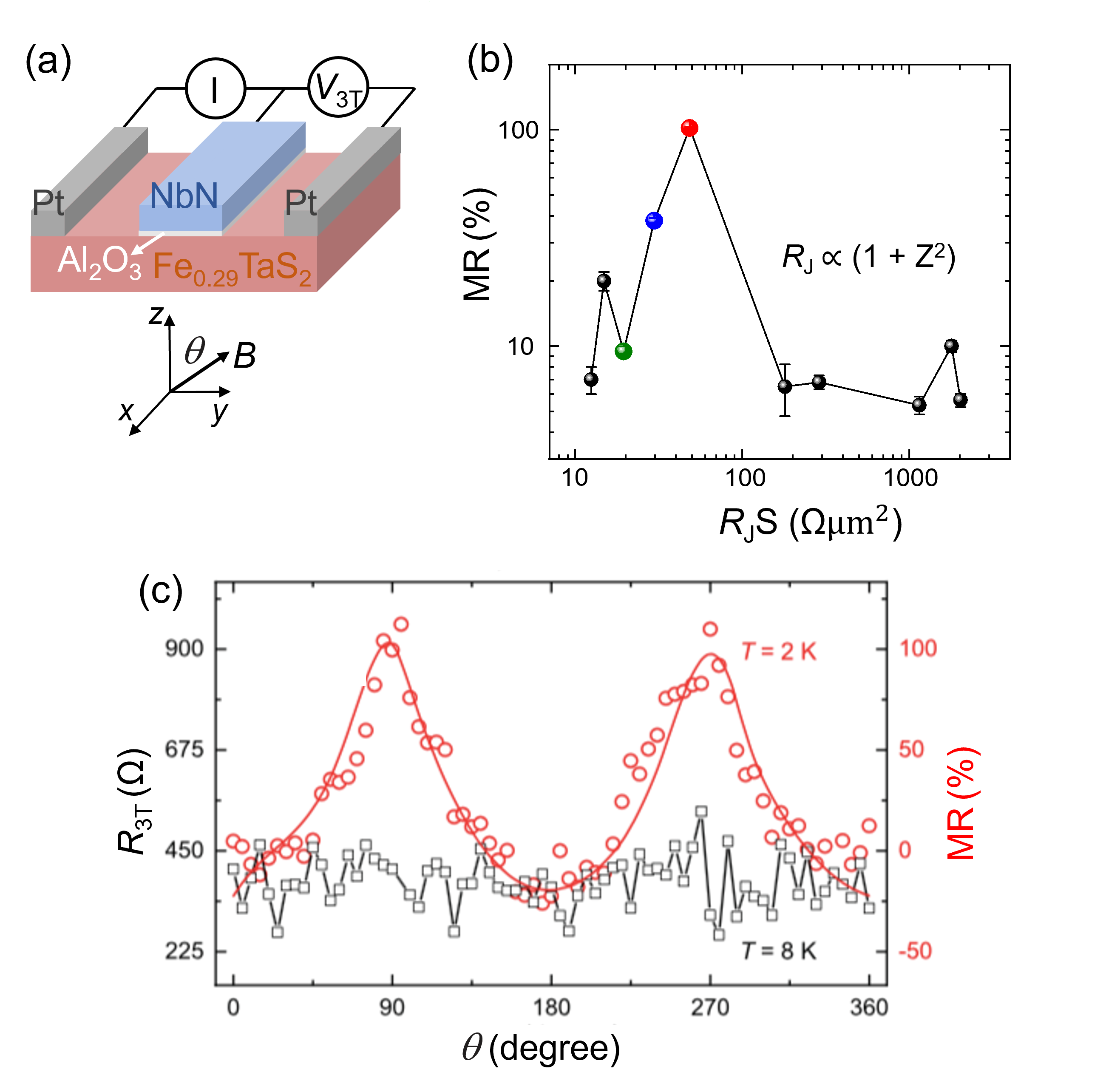}
	\caption{(a) Schematic of the experimental device, the superconductor NbN is separated from a ferromagnet Fe$_{0.29}$TaS$_2$ by a thin insulating Al$_2$O$_3$. Magnetic field, $B$, defined by the angle $\theta$, is applied in the $yz$ plane. (b) The low-bias amplitude of the magnetoresistance, MR, a relative difference between the three-terminal resistance, $R_{3T}=V_{3T}/I$ at $\theta=0$ and at $\theta=\pi/2$. The MR is a nonmonotonic in the resistance area product, $R_J S$, and therefore nonmonotonic in the interfacial barrier strength, $Z$, $R_J\propto (1+Z^2)$~\cite{Blonder1982:PRB}. (c) The angular dependence of MR at 2 K and $R_{3T}$ at 8 K. From Ref.~\onlinecite{Cai2021:NC}.}
	\label{fig:previous_work}
\end{figure}

Experimentally, interfacial magnetoanisotropy~\cite{Amundsen2024:RMP} can be used as a probe for such nonmonotonic trends in spin-triplet superconductivity.
One example is given in Fig.~\ref{fig:previous_work} from the nonmonotonic MR dependence on the resistance area product and, therefore, on the barrier
strength $Z$. Together with the observed angular and temperature dependence of the measured resistance, the corresponding 
magnetoanisotropy~\cite{Amundsen2024:RMP,Martinez2020:PRA,Cai2021:NC} is explained from the dominant contribution of the equal-spin Andreev reflection 
and spin-triplet superconductivity.

Remarkably, these trends with interfacial parameters are retained both within zero-bias conductance and, as we show in this work, from the correlation signatures of spin-triplet superconductivity.   
We confirm that these trends are present for both a simple step-function approximation of the pair potential and for the self-consistent pair potential.
Taken together, these findings are reassuring that the predicted large conductance magnetoanisotropy~\cite{Hogl2015:PRL,Vezin2020:PRB,Lv2018:PRB}, 
which can exceed by several orders of magnitude the corresponding normal-state values (with SOC and only one F region, such F/S junctions are good spin valves~\cite{Zutic2004:RMP}), 
offers a powerful experimental probe of the spin-triplet superconductivity~\cite{Martinez2020:PRA,Cai2021:NC}.

Following this introduction, in Sec.~II 
we describe employed methods to calculate conductance and superconducting correlation for F/S junctions.
In Sec.~III we explore 
nonmonotonic trends in spin-triplet superconducting pair correlations with interfacial parameters and 
compare them with the previously studied trends in conductance and the resulting magnetoanisotropy.
Our conclusions discuss implications of enhanced spin-triplet superconductivity, identified through complementary signatures of the conductance
and pair amplitudes, their experimental verification and possible future studies.  

\section{II. Methods}
\label{sec:Meth}
We consider a ballistic F/S junction depicted in Fig.~\ref{fig:scheme}. The system is assumed to be infinite in the plane parallel to the flat interface, which leads to the conservation of the parallel component of the wave vector, $\bm {k}_\|$, in all scattering processes. 
The F and S regions are assumed to be long enough for the bulk limit. We generalize the formalism of Griffin and Demers~\cite{Griffin1971:PRB} and Blonder-Tinkham-Klapwijk (BTK)~\cite{Blonder1982:PRB} to solve Bogoliubov-de Gennes equation, initially applied to normal metal/S and later to 
F/S junctions~\cite{Soulen1998:S,Zutic1999:PRBa,Zutic1999:PRBb,Kashiwaya1999:PRB,Zutic2000:PRB,Mazin2001:JAP,Kikuchi2001:PRB,Yamashita2003:PRB,Wu2009:PRB}. 
The two regions are separated by a $\delta$-function at $z=0$ to model the potential and SOC scattering at the interface.
For the $s$-wave superconducting pair potential, in addition to the usually
assumed step-function profile, $\Delta(z)=\Delta \Theta(z)$, we also provide its self-consistent solution obtained iteratively.
The corresponding  Bogoliubov-de Gennes equation for quasiparticle states ${\Psi_n}\left( \bm{r} \right) \equiv {e^{i{\bm{k_\parallel} } \cdot {\bm{r_\parallel} }}} \times {\left( {{u_{n \uparrow }}\left( z \right),{u_{n \downarrow }}\left( z \right),{v_{n \downarrow }}\left( z \right),{v_{n \uparrow }}\left( z \right)} \right)^T}$ with energy $E_n$ is
\begin{equation} \label{BdG equation}
\left( {\begin{array}{*{20}{c}}
	{{{\hat H}_e}}&{\Delta \left( z \right){I_{2 \times 2}}} \\ 
	{\Delta {{\left( z \right)}^*}{I_{2 \times 2}}}&{{{\hat H}_h}} 
	\end{array}} \right){\Psi_n}\left( \bm{r} \right) = {E_n}{\Psi_n}\left( \bm{r} \right),
\end{equation}
where the single-particle Hamiltonian for electron is 
\begin{equation} \label{H_e}
\begin{gathered}
{{\hat H}_e} =  - \frac{{{\hbar ^2}}}{2}\nabla \left[ {\frac{1}{{m\left( z \right)}}} \right]\nabla  - \mu \left( z \right) - \frac{{{\Delta _{xc}}}}{2}\bm{m} \cdot \bm{\hat \sigma }{\text{ }}\Theta \left( { - z} \right) \hfill \\
+ \left[ {{V_0}d + \alpha \left( {{k_y}{{\hat \sigma }_x} - {k_x}{{\hat \sigma }_y}} \right)} \right]\delta (z), \hfill \\ 
\end{gathered} 
\end{equation}
and for holes
\begin{equation} \label{H_h}
{{\hat H}_h} =  - {{\hat \sigma }_y}\hat H_e^*{{\hat \sigma }_y}.
\end{equation}
Here $m(z)$ is the effective mass, $\mu(z)$ the chemical potential, 
$\Delta_{xc}$ the exchange splitting in the ferromagnet, $\bm{m} = \left( {\sin \theta \cos  \phi , \sin \theta \sin \phi , \cos \theta } \right)$ 
is the magnetization orientation, defined by the spherical angles $\theta$ and $\phi$, 
while $\bm{\hat \sigma }$ is the vector of Pauli matrices $\hat\sigma_{x,y,z}$. 
F and S regions can have different effective masses, $m_{F,S}$ and Fermi wave vectors, $k_F, q_F$~\cite{Zutic1999:PRBa}, where $k_F$ is the spin-averaged value.
Generally, their mismatch in F and S region does not simply increase the effective interfacial barrier~\cite{Zutic1999:PRBa,Zutic2000:PRB}, 
as often assumed when extending the BTK approach~\cite{Blonder1983:PRB} to spin-polarized systems. 
For our analysis, we introduce the spin polarization $P={\Delta_{xc}}/{(2\mu_{F})}$
as well as dimensionless quantities to characterize the interfacial barrier and Rashba SOC strengths
\begin{eqnarray} 
Z &=&V_0 d \sqrt{m_F m_S}/(\hbar^2 \sqrt{k_F q_F}),  \\  
\lambda &=& 2\alpha \sqrt{m_F m_S}/ \hbar ^2. 
\label{eq:dimless}
\end{eqnarray}
To present trends for a large parameter space, we  
focus on the case $m_F=m_S=m$ and $k_F=q_F$,
where $k_F$ is the spin-averaged Fermi wave vector~\cite{Zutic1999:PRBa}. 

In the F region, the eigenspinors for electrons and holes can be written as 
$\chi _\sigma ^e = {\left( {{\chi _\sigma },0} \right)^T}$ and $\chi _\sigma ^h = {\left( {0,{\chi _{ - \sigma }}} \right)^T}$ with
\begin{equation}\label{spinor}
{\chi _\sigma } = {\left( {\sigma \sqrt {\frac{{1 + \sigma \cos \theta }}{2}} {e^{ - i\phi }},\sqrt {\frac{{1 - \sigma \cos \theta }}{2}} } \right)},
\end{equation}
where $\sigma=1 (-1)$ refer to spin parallel (antiparallel) to $\bm{M}$.
As in the BTK formalism~\cite{Blonder1982:PRB}, the wave functions in the F and S regions can be expressed as a linear combination of all possible eigenstates.
As expected from the Snell's law~\cite{Zutic2000:PRB}, for a large $k_\parallel$, $z$-components of the wave vectors in the F region, $k^{e\,(h)}_\sigma = \sqrt{k_F^2+(2m_F/\hbar^2)\left[(-)E + \sigma \Delta_{xc}/2\right]-k_\parallel^2}$, 
can become imaginary representing evanescent states which carry no net current~\cite{Zutic2000:PRB,Zutic1999:PRBa,Kashiwaya1999:PRB}.
With the matching of the wave functions in the F and S regions, from the charge current conservation, we can express zero-temperature conductance at applied bias, $V$,
\begin{equation}
G(V) =  \sum\limits_\sigma  \int \frac{d \bm{k_\parallel}}{2 \pi k_F^2} \left[ 1 + R_\sigma^h(eV) - R_\sigma^e(eV) \right],
\label{eq:G}
\end{equation}
normalized by the Sharvin conductance $G_\mathrm{Sh} = e^2 k_F^2 A/(2 \pi h)$~\cite{Zutic2004:RMP}, where $A$ is the interfacial area.
Only the probability amplitudes from the F region are needed for  Andreev  $R_\sigma^h$ and specular reflection $R_\sigma^e$, which contain both processes with and without interfacial spin-flip  scattering~\cite{Zutic1999:PRBa,Hogl2015:PRL,Vezin2020:PRB}.

To numerically study the F/S system from Fig.~\ref{fig:scheme}, we divide the superconducting regions into multiple thin layers with thicknesses $~k_F^{-1}$ and treat  
$\Delta(z)$ as a constant through a single layer. The wave functions in each layer are also constructed from the linear combinations of all possible eigenstates in that layer. All the boundary conditions connecting the wave functions in adjacent layers form a system of linear equations that can be solved using the transfer matrix method \cite{Wu2012:PRB}. The wave functions can thus be obtained from the solutions of the linear equations. 
The self-consistent pair potential can be expressed in terms of the wave functions
as~\cite{Simensen2018:PRB,Valls2010:PRB,Beiranvand2016:PRB,Wu2012:PRB,Halterman2013:PRL,Halterman2007:PRL,Halterman2008:PRB,Halterman2001:PRB,Halterman2002:PRB,Alidoust2015:PRB}, where we do not assume the quasiclassical approximation~\cite{Amundsen2024:RMP,Eschrig2015:RPP,Belzig1999:SM}.
It is useful to write such a self-consistent solution as $\Delta \left( z \right) = g\left( z \right)F\left( z \right)$,
where $g(z)$ is the superconducting coupling constant that is a constant in the S region and is zero in the F region, while $F(z)$ is the Cooper pair amplitude~\cite{Valls2010:PRB}, which we write by choosing the spin-quantization axis along $\bm{z}$
\begin{equation}
\begin{gathered}
F\left( z \right) = \frac{1}{2}\left\langle {{\psi _ \uparrow }\left( {z,0} \right){\psi _ \downarrow }\left( {z,0} \right) - {\psi _ \downarrow }\left( {z,0} \right){\psi _ \uparrow }\left( {z,0} \right)} \right\rangle  \hfill \\
= \frac{1}{2}\sum\limits_n {\left[ {{u_{n \uparrow }}\left( z \right)v_{n \downarrow }^*\left( z \right) + {u_{n \downarrow }}\left( z \right)v_{n \uparrow }^*\left( z \right)} \right]}  \hfill \\
\times \tanh \left( {{E_n}/2{k_B}T} \right), \hfill \\ 
\label{eq:singlet}
\end{gathered}
\end{equation}
where ${\psi _ \sigma}$ is the annihilation operator for electron with spin $\sigma$, evaluated at time zero, $k_B$ is the Boltzmann constant and $T$ is the temperature. The sum is over all possible eigenstates, with the excitation range set by the Debye energy, $E_n < \hbar \omega_D \approx 0.1 \mu_F$.

We start from an initial guess $\Delta(z)$ that we chose to be a step-function  profile 
and solve the linear equations described above to obtain the new pair potential, which will be used in the next iteration. This process is repeated until the pair potential converges. Once the self-consistent pair potential and wave functions are obtained, we calculate the time-dependent triplet-pair amplitudes corresponding to the spin projection $s_z = 0$, $s_z =\pm 1$ by, respectively~\cite{Simensen2018:PRB,Valls2010:PRB,Beiranvand2016:PRB,Wu2012:PRB,Halterman2013:PRL,Halterman2007:PRL,Halterman2008:PRB,Halterman2001:PRB,Halterman2002:PRB,Alidoust2015:PRB}
\begin{equation}\label{f0}
\begin{gathered}
{f_0}\left( {z,t} \right) = \frac{1}{2}\left\langle {{\psi _ \uparrow }\left( {z,t} \right){\psi _ \downarrow }\left( {z,0} \right) + {\psi _ \downarrow }\left( {z,t} \right){\psi _ \uparrow }\left( {z,0} \right)} \right\rangle  \\ 
= \frac{1}{2}\sum\limits_n {\left[ {{u_{n \uparrow }}\left( z \right)v_{n \downarrow }^*\left( z \right) - {u_{n \downarrow }}\left( z \right)v_{n \uparrow }^*\left( z \right)} \right]} {\zeta  _n}\left( t \right), \\ 
\end{gathered}
\end{equation}
\vspace{-0.6cm}
\begin{equation}\label{f1}
\begin{gathered}
{f_1}\left( {z,t} \right) = \frac{1}{2}\left\langle {{\psi _ \uparrow }\left( {z,t} \right){\psi _ \uparrow }\left( {z,0} \right) - {\psi _ \downarrow }\left( {z,t} \right){\psi _ \downarrow }\left( {z,0} \right)} \right\rangle  \\ 
=  - \frac{1}{2}\sum\limits_n {\left[ {{u_{n \uparrow }}\left( z \right)v_{n \uparrow }^*\left( z \right) + {u_{n \downarrow }}\left( z \right)v_{n \downarrow }^*\left( z \right)} \right]} {\zeta _n}\left( t \right), \\ 
\end{gathered}
\end{equation}
where the time-dependent function is
${\zeta _n}\left( t \right) = \cos \left( {{E_n}t/\hbar } \right) - i\sin \left( {{E_n}t/\hbar } \right)\tanh \left( {{E_n}/2{k_B}T} \right)$.

\section{III. Results}
\label{sec:Res}
\begin{figure}[t]
	\centering
	\includegraphics*[trim=-1.8cm 3.5cm 0.cm 2.3cm,clip,width=15.5cm]{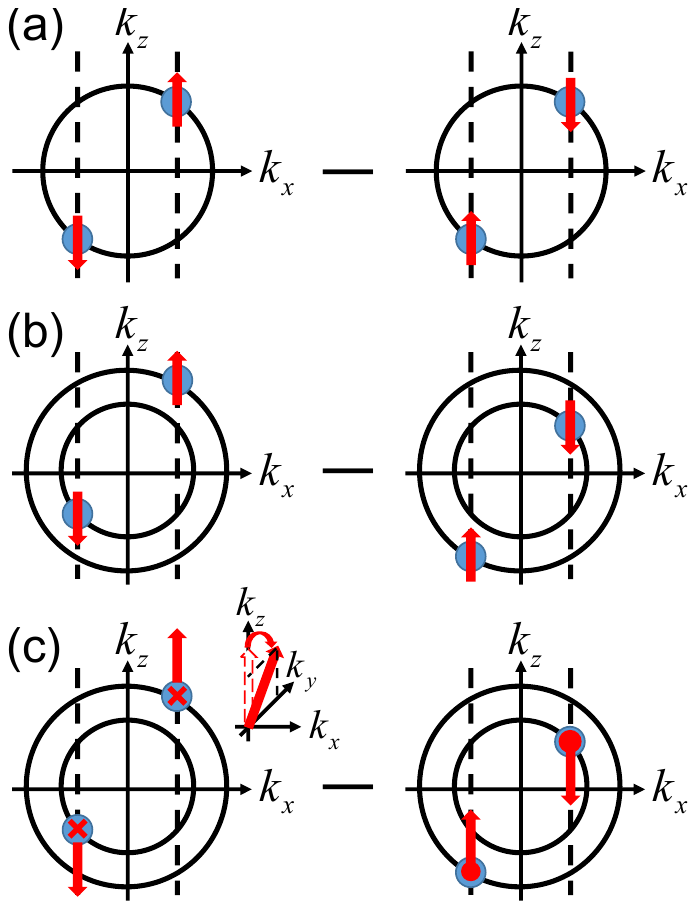}
	\caption{Evolution of a spin-singlet Cooper pair under ferromagnetism and SOC. For simplicity, we take $k_y=0$ and an out-of-plane ${\bm M}$. (a) Spin-singlet Cooper pair in an $s$-wave superconductor. (b) The FFLO state is formed when the singlet Cooper pair enters F through Andreev reflection. The equi-energy contours split for spin up and down due to ${\bm M}$. The in-plane $k_\|$, parallel to the interface is preserved. (c) The spins in an FFLO state are tilted by the interfacial Rashba SOC in an F/S junction. The spins  are not completely antiparallel in a pair and thus the equal-spin pairing arises.}
	\label{fig:pert}
\end{figure}

The spin-triplet paring arises from interplay between the magnetism and SOC, transforming the spin-singlet pairs in an $s$-wave superconductor. Such a process occurs at the F/S interface, and the emergent spin-triplet pair penetrates both into the F and S region due to the proximity effect.
Recalling Fig.~\ref{fig:scheme}(a) that $z$-direction is perpendicular to the interface and focusing on the $k_y=0$ case, we sketch the evolution of the $s$-wave Cooper 
pair first by showing it in the absence of both $\bm{M}$ and SOC in Fig.~\ref{fig:pert}(a).
We consider an out-of-plane $\bm{M}$, which spin-splits the Fermi surface as shown in Fig.~\ref{fig:pert}(b). Due to the conservation of energy and the interfacial wave vector, the $k_z$ values on the Fermi surface are shifted from those in a singlet Cooper pair in Fig.~\ref{fig:pert}(a).
For clarity, the conservation of $k_\|$ is denoted by dashed vertical lines in Fig.~\ref{fig:pert}.
Consequently, the Cooper pair in the F region acquires a center-of-mass momentum, which leads 
to the decaying pair-amplitude oscillations, consistent with the understanding of the Fulde-Ferrell-Larkin-Ovchinnikov (FFLO) state~\cite{Fulde1964:PR,Larkin1965:SPJETP}.
In the limit of $\Delta / \mu \to 0$,
the center-of-mass momentum acquired by $\left| {k \uparrow , - k \downarrow } \right\rangle$ along $k_z$ is $\hbar Q \approx {\Delta _{xc}}/ {2 v_{Fz}}$ [left panel in Fig.~\ref{fig:pert}(b)], where $v_{Fz}$ is the $z$-component of the Fermi velocity,
while $\left| k  \downarrow , - k \uparrow \right\rangle$ acquires $-\hbar Q$. 
Due to the exchange splitting, the singlet state $\left| {k \uparrow , - k \downarrow } \right\rangle  - \left| {k \downarrow , - k \uparrow } \right\rangle$ becomes (when $\bm{M}$ is out-of-plane)~\cite{Eschrig2015:RPP}
\begin{equation}\label{psi_OP}
	\left| {{\psi _{OP}}} \right\rangle  = \left| {k + Q \uparrow , - k + Q \downarrow } \right\rangle  - \left| {k - Q \downarrow , - k - Q \uparrow } \right\rangle
\end{equation}

To distinguish the pairing properties of the states, we define the pairing symmetries states for spin-singlet ($s$), $s_z=0$ triplet ($f_0$), and $s_z=\pm 1$ triplet ($f_1$) as
\begin{equation}\label{sym_state}
	\begin{gathered}
		\left| {P\left( s \right)} \right\rangle  = \frac{1}{2}\left( {\left| {{z_1} \uparrow ,{z_2} \downarrow } \right\rangle  - \left| {{z_1} \downarrow ,{z_2} \uparrow } \right\rangle } \right), \hfill \\
		\left| {P\left( {{f_0}} \right)} \right\rangle  = \frac{1}{2}\left( {\left| {{z_1} \uparrow ,{z_2} \downarrow } \right\rangle  + \left| {{z_1} \downarrow ,{z_2} \uparrow } \right\rangle } \right), \hfill \\
		\left| {P\left( {{f_1}} \right)} \right\rangle  = \frac{1}{2}\left( {\left| {{z_1} \uparrow ,{z_2} \uparrow } \right\rangle  - \left| {{z_1} \downarrow ,{z_2} \downarrow } \right\rangle } \right), \hfill \\ 
	\end{gathered}
\end{equation}
where the positions are taken at $x = y = 0$ and at a general $z$.
Taking the projection of $\left| {{\psi _{OP}}} \right\rangle$ onto $\left|P\right\rangle$, and taking $z_1, z_2$ to be close to each other ($z_1,z_2 \to z$), we find: $\braket{P\left( s \right)|\psi_{OP}} = \cos \left( {2Qz} \right)$, $\braket{P\left( f_0 \right)|\psi_{OP}} = i \sin \left( {2Qz} \right)$ and $\braket{P\left( f_1 \right)|\psi_{OP}} = 0$.
This indicates that the unperturbed $\left| {{\psi _{OP}}} \right\rangle$ contains both the spin-singlet $s$ and the spin-triplet $f_0$ components.

We note that the analysis above assumes the spin-quantization axis along $\bm{M}$, which describes the out-of-plane $\bm{M}$ in our system.
However, for an in-plane $\bm{M} \| \bm{x}$, we construct the mixed-spin state onto the spin-quantization axis along $\bm{x}$
\begin{equation}\label{psi_IP}
	\begin{gathered}
		\left| {{\psi _{IP}}} \right\rangle  = \left( {\left| {k + Q{ \uparrow _x}, - k + Q{ \downarrow _x}} \right\rangle  - \left| {k - Q{ \downarrow _x}, - k - Q{ \uparrow _x}} \right\rangle } \right), \hfill \\
	\end{gathered}
\end{equation} 
where $\uparrow_x, \downarrow_x$ denotes the spin states parallel and antiparallel to $\bm{x}$, respectively. Projecting this state onto the spin-quantization axis along $\bm{z}$ (Appendix B), we get $\braket{P\left( s \right)|\psi_{IP}} = \cos \left( {2Qz} \right)$, $\braket{P\left( f_0 \right)|\psi_{IP}} = 0$ and $\braket{P\left( f_1 \right)|\psi_{IP}} = - i \sin \left( {2Qz} \right)$.
This shows that $f_1$ triplet emerges for an in-plane $\bm{M}$ in the F/S junction even without SOC.
Such rotation of spin-quantization axis corresponds to the situation of noncollinear $\bm{M}$ in fabricated junctions~\cite{Eschrig2015:RPP,Singh2015:PRX}.

With interfacial SOC in the F/S junction, spin-flip process during Andreev reflections~\cite{Vezin2020:PRB} allows the formation of equal-spin pairing.
In Fig.~\ref{fig:pert}(c), we see that the spins in the FFLO state are tilted by the SOC, which leads to the emergence of the equal-spin component in a Cooper pair and can influence the superconducting transition temperature~\cite{Banerjee2018:PRB}.
To further understand this spin tilting by SOC, we treat the interfacial SOC as a perturbation of the mixed-spin 
state in the F/S junction [Fig.~\ref{fig:pert}(b)]. It can be shown (see Appendix B) that, for an out-of-plane $\bm{M}$, the projections in the first order of the perturbed state $\ket{\psi _{OP}^{\left( 1 \right)}}$ onto $\left|P\right\rangle$ are $\braket{P\left( s \right)|\psi_{OP}^{\left( 1 \right)}} = 0$,
$\braket{P\left( f_0 \right)|\psi_{OP}^{\left( 1 \right)}} = 0$ and 
$\braket{P\left( f_1 \right)|\psi_{OP}^{\left( 1 \right)}} \propto - 2\alpha {k_y}\cos \left( {2Qz} \right)$.
Thus the $f_1$ triplet arises when Rashba SOC is applied to the F/S interface when $\bm{M}$ is out-of-plane.

On the other hand, for an in-plane $\bm{M} \| \bm{x}$, these projections become (see Appendix B) $\braket{P\left( s \right)|\psi_{IP}^{\left( 1 \right)}} \propto 2i\alpha {k_y}\sin \left( {2Qz} \right)$,
$\braket{P\left( f_0 \right)|\psi_{IP}^{\left( 1 \right)}} = 0$ and 
$\braket{P\left( f_1 \right)|\psi_{IP}^{\left( 1 \right)}} \propto - 2\alpha {k_y}\cos \left( {2Qz} \right)$, which also contain $f_1$ triplet. 
 
\begin{figure}[b]
	\centering
	\includegraphics*[trim=0.4cm 3.2cm 0.4cm 2.9cm,clip,width=13.5cm]{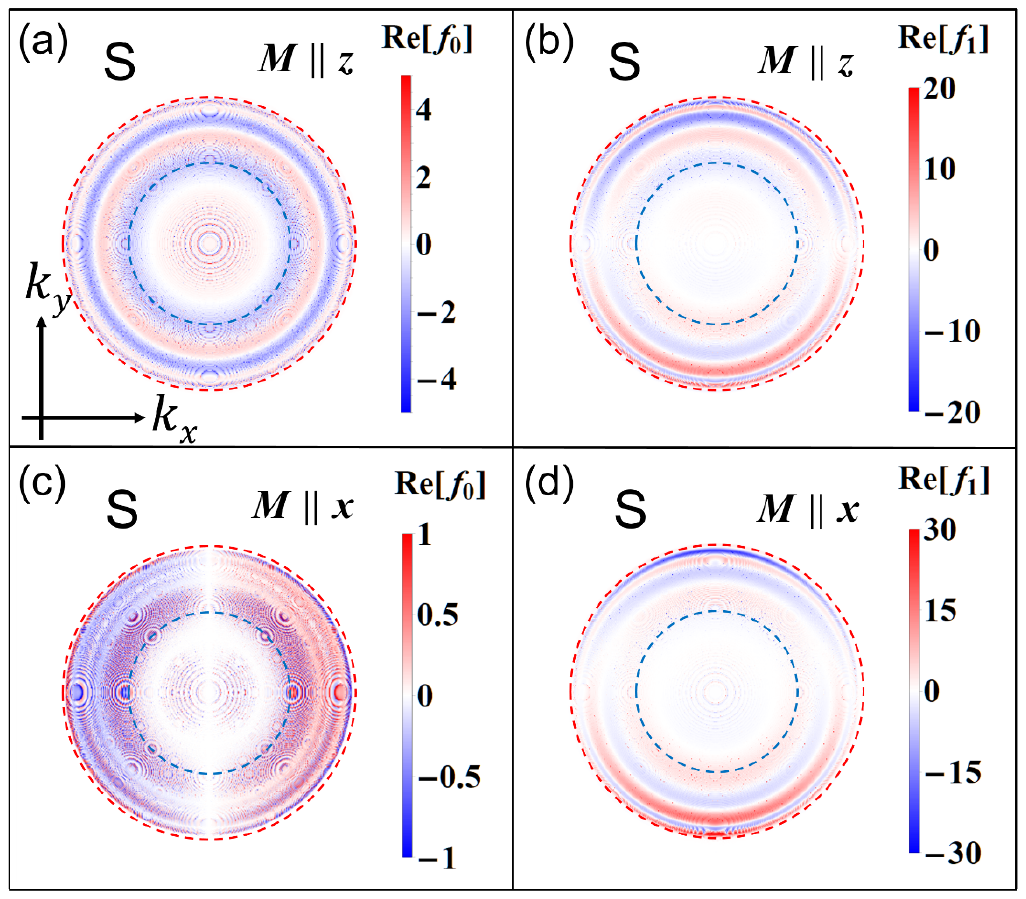}
	\caption{$\bm{k_\parallel}$-resolved $\text{Re}[f_0]$ (left column) and  $\text{Re}[f_1]$ (right column) at fixed position $z=10k_F^{-1}$ in the S region. The top (bottom) row is for an out-of-plane (in-plane) ${\bm M}$. The red (blue) circle has a radius of the spin-averaged (spin-down) Fermi wave vector.}
	\label{fig:kp_S}
\end{figure}

\begin{figure}[t]
	\centering
	\includegraphics*[trim=0.5cm 3.4cm 0.4cm 2.9cm,clip,width=13.5cm]{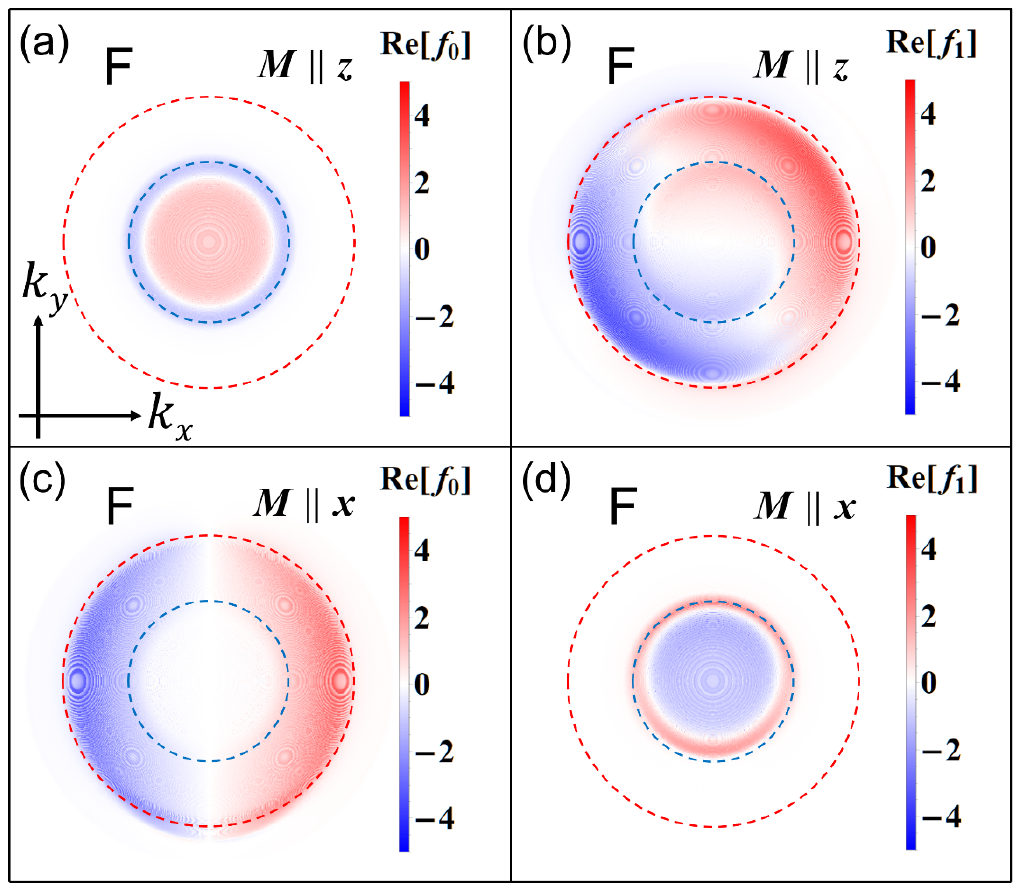}
	\caption{$\bm{k_\parallel}$-resolved $\text{Re}[f_0]$ (left column) and  $\text{Re}[f_1]$ (right column) at fixed position $z=-10k_F^{-1}$  in the F region. The top (bottom) row is for an out-of-plane (in-plane)  ${\bm M}$. The red (blue) dashed circle has a radius of the spin-averaged (spin-down) Fermi wave vector.}
	\label{fig:kp_F}
\end{figure}

These conclusions derived from the simplified toy model can help us understand the pairing symmetry in F/S junctions with interfacial SOC and provide guidance for the correlations calculated for the S and F region, shown in Figs.~\ref{fig:kp_S} and \ref{fig:kp_F}.
In Fig.~\ref{fig:kp_S}(a) the calculated $\bm{k_\parallel}$-resolved $f_0$ triplet shows an even parity for $\bm{M} \| \bm{z}$, which corresponds to $\braket{P\left( f_0 \right)|\psi_{OP}} = i \sin \left( {2Qz} \right)$
from the unperturbed state $\left| {{\psi _{OP}}} \right\rangle$.
In contrast, the $f_1$ triplet has an antisymmetric pattern for both out-of-plane and in-plane $\bm{M}$ in Figs.~\ref{fig:kp_S}(b) and \ref{fig:kp_S}(d).
This agrees with the odd parity with $k_y$ in the projections onto $f_1$: $\braket{P\left( f_1 \right)|\psi_{OP}^{\left( 1 \right)}} \propto - 2\alpha {k_y}\cos \left( {2Qz} \right)$, $\braket{P\left( f_1 \right)|\psi_{IP}^{\left( 1 \right)}} \propto - 2\alpha {k_y}\cos \left( {2Qz} \right)$.
While for an in-plane $\bm{M}$ in Fig.~\ref{fig:kp_S}(c) we see an antisymmetric pattern even for $f_0$ triplet, it  is related to the higher-order perturbation, 
as its magnitude is much smaller than the remaining results in Fig.~\ref{fig:kp_S}.
 
In the F region, the $\bm{k_\parallel}$-resolved correlation functions shown in Fig.~\ref{fig:kp_F} follow the 
similar trends as those in the S region, 
but the pattern is more complicated due to the spin-precession across the whole region.
Nevertheless, to understand the pair correlations in F/S junctions with SOC, we can already
gain some some valuable guidance from the prior analysis without SOC~\cite{Eschrig2015:RPP,Demler1997:PRB}.

\begin{figure}[t]
	\centering
	\includegraphics*[trim=0.35cm 0.6cm 0cm 0.9cm,clip,width=8.75cm]{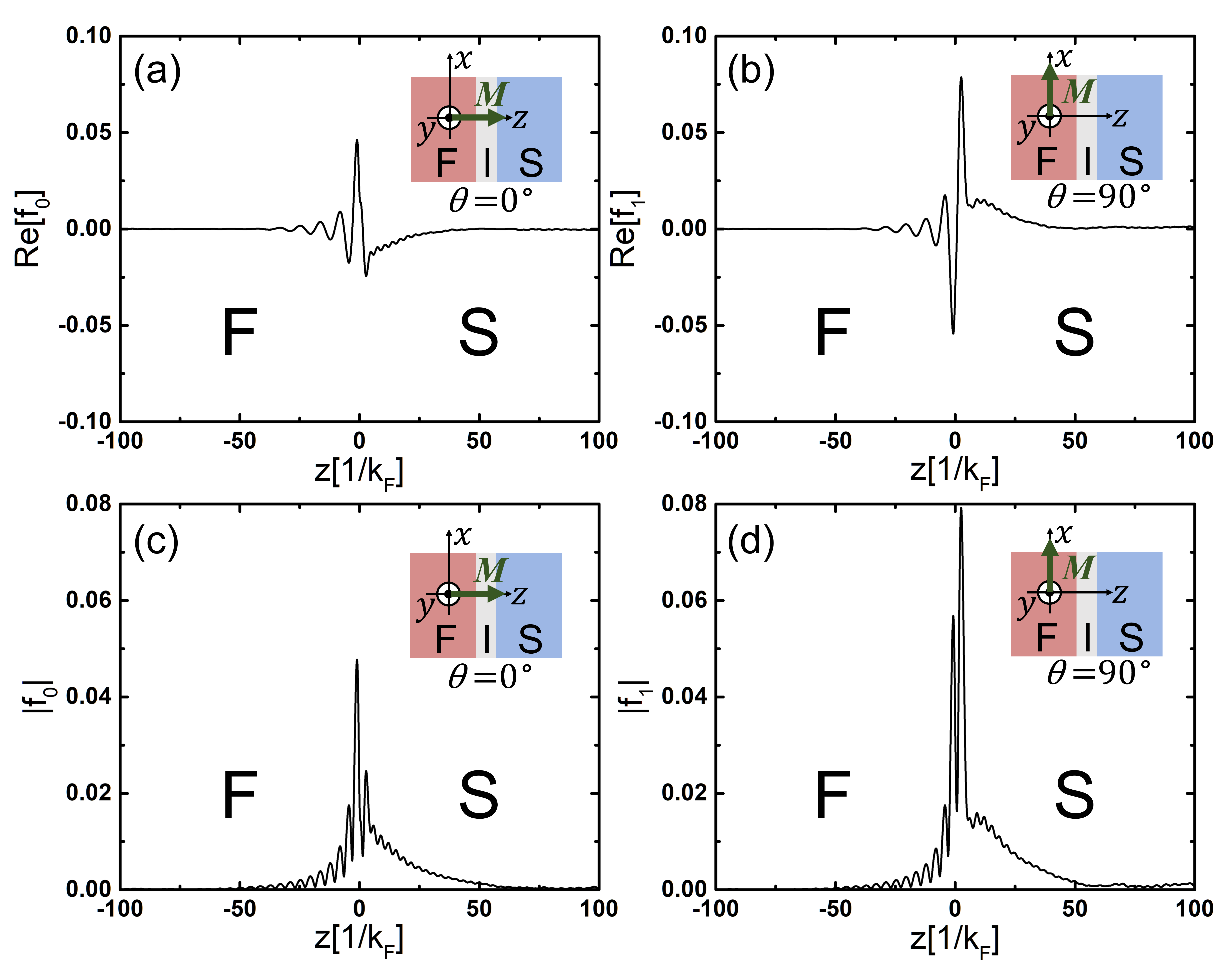}
	\caption{(a), (c) Spatial dependence of $s_z = 0$ triplet-pair amplitudes for $Z=2$, $\lambda=5$, $\tau = 4$ and an out-of-plane $\bm{M}$ with $P = 0.7$. (b), (d) The same plots for $s_z = \pm 1$ triplet-pair amplitudes with an in-plane $\bm{M}$. } 
	\label{fig:triplet_in_out}
\end{figure}

We next investigate the spatial information of the spin-triplet pair amplitudes $f_0$, $f_1$, which are directly calculated from Eqs.~(\ref{f0}) and (\ref{f1}). We show in the Appendix A that the self-consistency can be achieved in just a few iterations, while the results from the step-function profile $\Delta(z)$ can already capture the main features.
Therefore, we will mostly focus on the results from a step-function $\Delta(z)$. We also consider the low-temperature regime and take Debye energy $\hbar \omega_D = 0.1 \mu_F$. The BCS coherence length is taken as $\xi_0  = 50 k^{-1}_F$ and all the correlation functions are normalized to the singlet-pair 
amplitude inside the bulk. We define dimensionless time as $\tau \equiv \omega_D t$.

We first consider an out-of-plane and in-plane $\bm{M}$, where the calculated pair amplitudes at a fixed $Z$ and $\lambda$ can be further analyzed from their underlying symmetries discussed above.
The nonvanishing triplet amplitudes for these two cases are shown in Fig.~\ref{fig:triplet_in_out}, where we again choose $P=0.7$. This value of spin polarization
is further motivated by the experiments from Ref.~\cite{Martinez2020:PRA} where in all-epitaxial Fe/MgO/V junctions both an out-of-plane and in-plane $\bm{M}$ are stable states, even without an external applied magnetic field.

From Fig.~\ref{fig:triplet_in_out}, we see that for an out-of-plane $\bm{M}$ (in-plane $\bm{M}$), $f_1$  ($f_0$) vanishes, 
while in the each considered case there is a spatial 
oscillatory decay of the nonvanishing components, 
away from the interface into the F and S region.
These results are in agreement with the analysis above and the previous study~\cite{Simensen2018:PRB}. 
As is discussed above, the vanishing of these triplet-pair amplitudes is due to the antisymmetric correlation functions in the $k_{\parallel}$ plane (Figs.~\ref{fig:kp_S} and \ref{fig:kp_F}). 
For an out-of-plane $\bm{M}$ (in-plane $\bm{M}$), $f_1$ ($f_0$) is an antisymmetric function in the $k_{\parallel}$ plane, along the F/S interface.
In the absence of SOC, the evolution of pair correlations in F/S junctions with exchange splitting is well understood~\cite{Eschrig2015:RPP,Demler1997:PRB}.

\begin{figure}[!htb]
	\vspace{-0.2cm}
	\centering
	\includegraphics*[trim=0.9cm 0cm 0cm 0cm,clip,width=8.75cm]{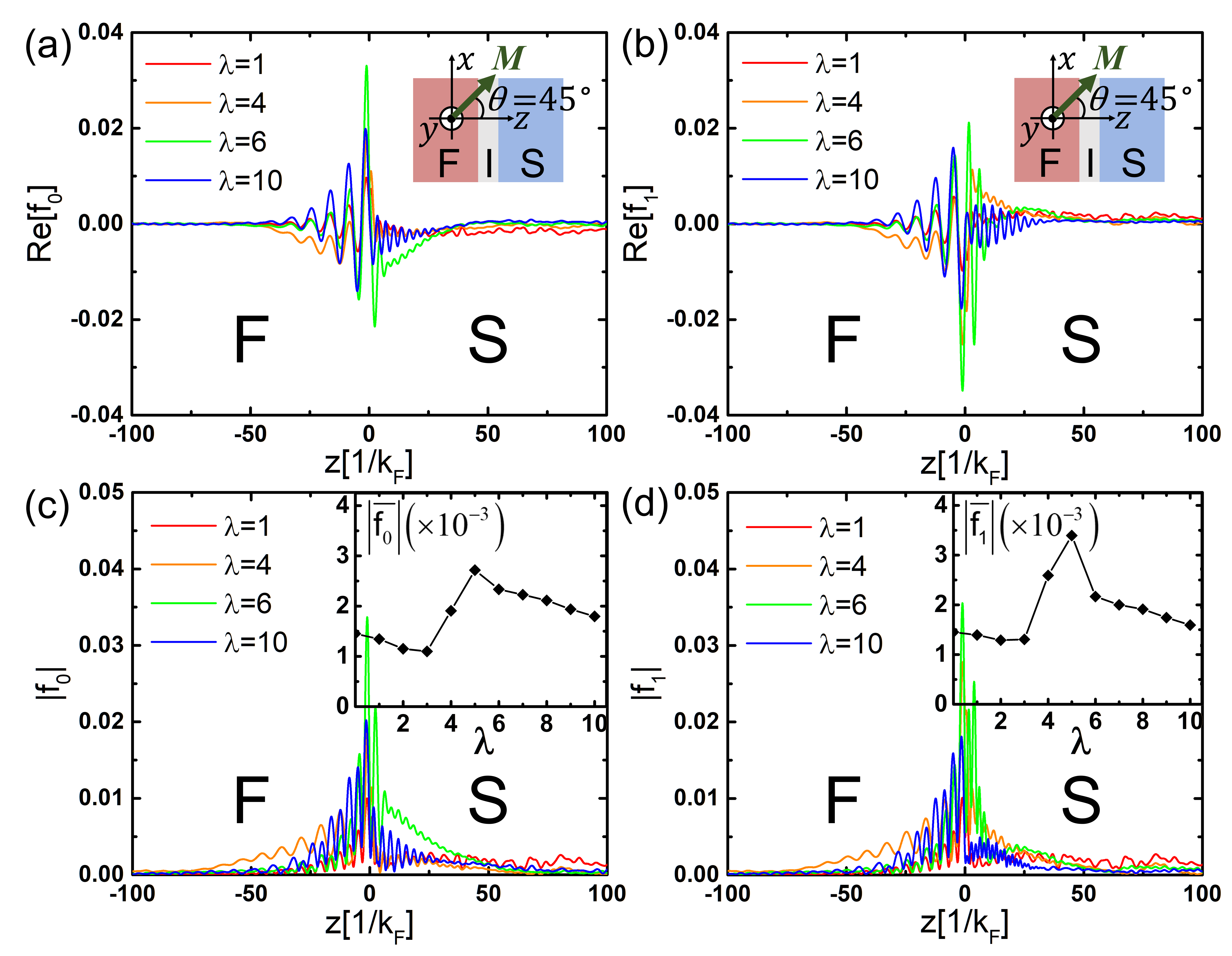}
	\caption{(a), (c) Spatial dependence of the real part and absolute value of the $s_z = 0$ triplet-pair amplitudes for different SOC strength $\lambda$,
		when $Z=2$, $\tau = 4$, and $P = 0.7$. (b), (d)	The same plots for $s_z = \pm 1$ triplet-pair amplitudes. The color codes are the same in (a)-(d).
		The insets of (c), (d) show the spatially-averaged absolute value of the triplet-pair amplitudes as a function of $\lambda$. The interface is at $z=0$.}
	\label{fig:triplet_fixing Z}
\end{figure}

To study the impact of interfacial parameters on the nonvanishing spin-triplet pair amplitudes, 
we choose a general $\bm{M}$ with a polar angle $\theta = \pi / 4$, while the azimuthal angle can be arbitrary 
due to the system's rotational symmetry with respect to $z$-axis.
We first examine the role of the SOC strength on the pair amplitudes in Fig.~\ref{fig:triplet_fixing Z} at a fixed $Z=2$. 
From Figs.~\ref{fig:triplet_fixing Z}(a) and 
\ref{fig:triplet_fixing Z}(b), we can see that the correlations for $s_z = 0$ and $s_z = \pm 1$ triplet 
pairing arise at the interface and their real part undergoes decaying oscillations in both F and S. 
The oscillation period is the same for different $\lambda$.  
The spatial extent of the correlations is close to the BCS coherence length. 
Since $f_0$ and $f_1$ are generally complex, in Figs.~\ref{fig:triplet_fixing Z}(c) and \ref{fig:triplet_fixing Z}(d) we also show
their absolute value. 

Instead of the usual spatially-resolved pair amplitudes, in the insets of Figs.~\ref{fig:triplet_fixing Z}(c) and 
\ref{fig:triplet_fixing Z}(d) we give the spatially-averaged values 
of $|f_0|$ and $|f_1|$. This offers a better comparison with the previously analyzed zero-bias conductance~\cite{Vezin2020:PRB}
and the measured MR in Fig.~\ref{fig:previous_work} which, similarly, 
have no spatial information. 
Surprisingly, this pair-amplitude behavior, in addition to being nonmonotonic in $\lambda$, has a peak around $\lambda=5$
and close to the maximum position of the triplet-related component of $G$ due to the spin-flip Andreev reflection.
The nonmonotonic behavior of the spin-flip Andreev reflection arises
from the effective barrier strength~\cite{Vezin2020:PRB} 
\begin{equation}
	Z^\pm_{\rm{eff}} = Z \pm {\lambda}{k_\parallel } /(2\sqrt{{k_F}{q_F}}), 
	\label{eq:Zeff_corr}
\end{equation}
where $Z^+_{\rm{eff}}$ ($Z^-_{\rm{eff}}$) is for inner (outer) Rashba bands [see Fig.~\ref{fig:scheme}(b)]. When $Z\geq 0$ and $\lambda \geq 0$, $Z^+_{\rm{eff}} \geq Z$,
the effective barrier is increased. 
However, at open channels $k_\parallel= (2Z/\lambda)\sqrt{k_F q_F}$, $Z^-_{\rm{eff}}=0$ and gives a dramatically increased $G$. 
The maximum $G$ is achieved when the amount of the open channels
$\propto k_\parallel$, is maximized. Therefore, the maximum spin-flip Andreev reflection is located 
near $q_F= (2Z/\lambda)\sqrt{k_F q_F}$, i.e., $\lambda=2Z$ when $k_F = q_F$.
This agreement of the maximum conditions between the spin-flip Andreev reflection and the triplet-pair amplitude suggests that the 
enhancement mechanism proposed in Ref.~\cite{Vezin2020:PRB} also applies to the triplet-pair amplitude. 

\begin{figure}[t]
	\centering
	\includegraphics*[trim=0.9cm 0.9cm 0cm 0.2cm,clip,width=8.75cm]{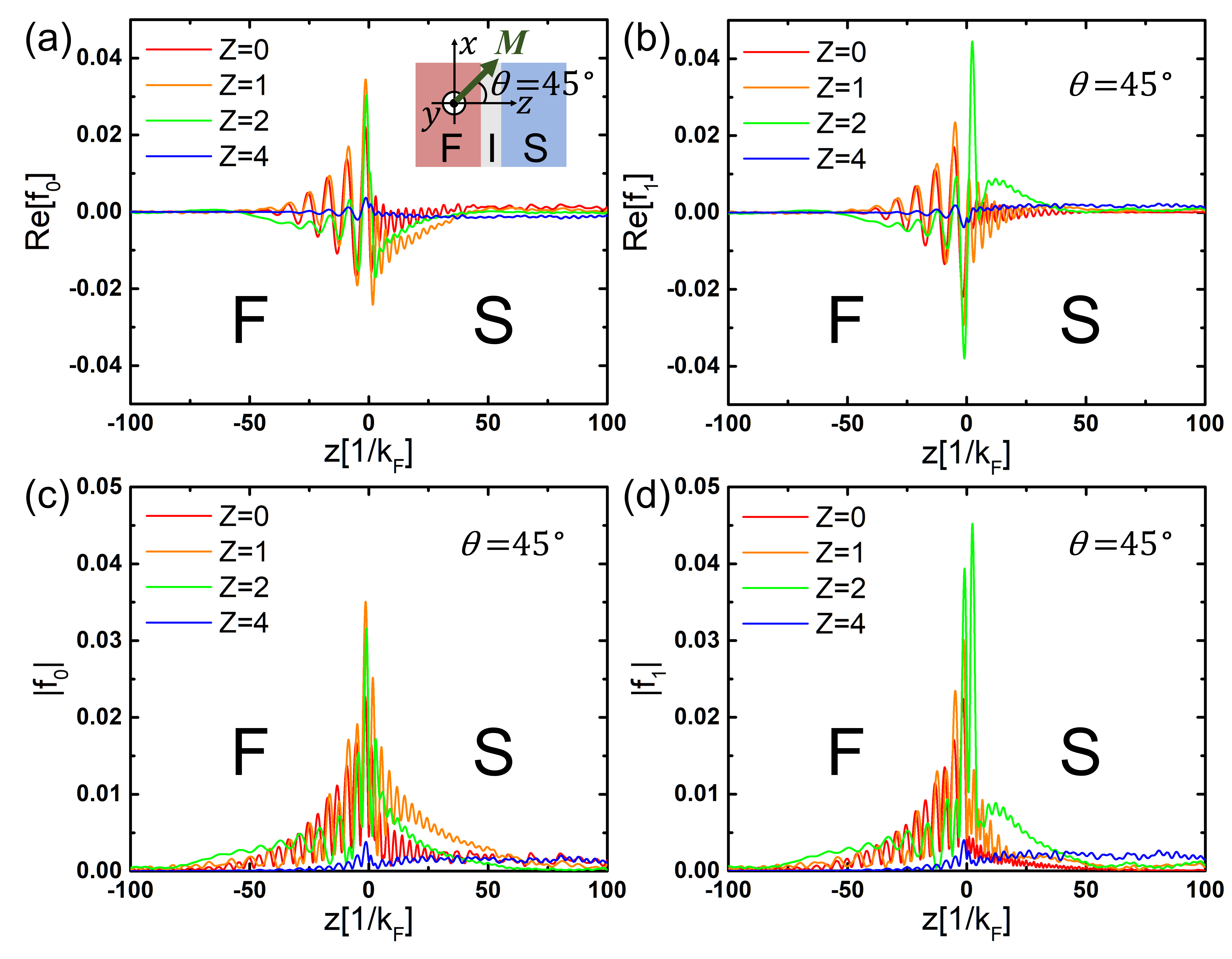}
	\caption{(a), (c) Spatial dependence of the real part  and absolute value of the $s_z = 0$ triplet-pair amplitudes for different barrier potential $Z$ when $\lambda=5$, $\tau = 4$ and $P = 0.7$. (b), (d) The same plots for $s_z = \pm 1$ triplet-pair amplitudes.} 
	\label{fig:triplet_fixing la}
\end{figure}

We next consider the role of barrier strength on $f_0$ and $f_1$, in Fig.~\ref{fig:triplet_fixing la} at a fixed $\lambda=5$. 
Similar as in Figs.~\ref{fig:triplet_fixing Z}(a) and \ref{fig:triplet_fixing Z}(b), we see that such decaying oscillations 
also appear in Figs.~\ref{fig:triplet_fixing la}(a) and \ref{fig:triplet_fixing la}(b) when SOC is fixed.
If we just focus on the F region, we find another similarity with Fig.~\ref{fig:triplet_fixing Z}, both the real parts and the absolute values of $f_0$ and $f_1$ appear nearly identical.
A large $Z$ can strongly suppress both $f_0$ and $f_1$, but a completely transparent barrier ($Z=0$) does not maximize the triplet-pair amplitude. Both $f_0$ and $f_1$ are nonmonotonic with respect to $Z$, and they peak around $Z=1$. 

Taken together, these results from Figs.~\ref{fig:triplet_fixing Z} and ~\ref{fig:triplet_fixing la} confirm the conclusions made in Ref.~\cite{Vezin2020:PRB}: 
Instead of a stronger SOC or a weaker potential barrier, a suitable match of $Z$ and $\lambda$ supports the 
strongly-enhanced triplet pairing in our system. 
Thus, such a condition enhances both triplet current and triplet-proximity effects.

\begin{figure}[t]
	\centering
	\includegraphics*[trim=1.0cm 0.2cm 0.6cm 0.8cm,clip,width=8.75cm]{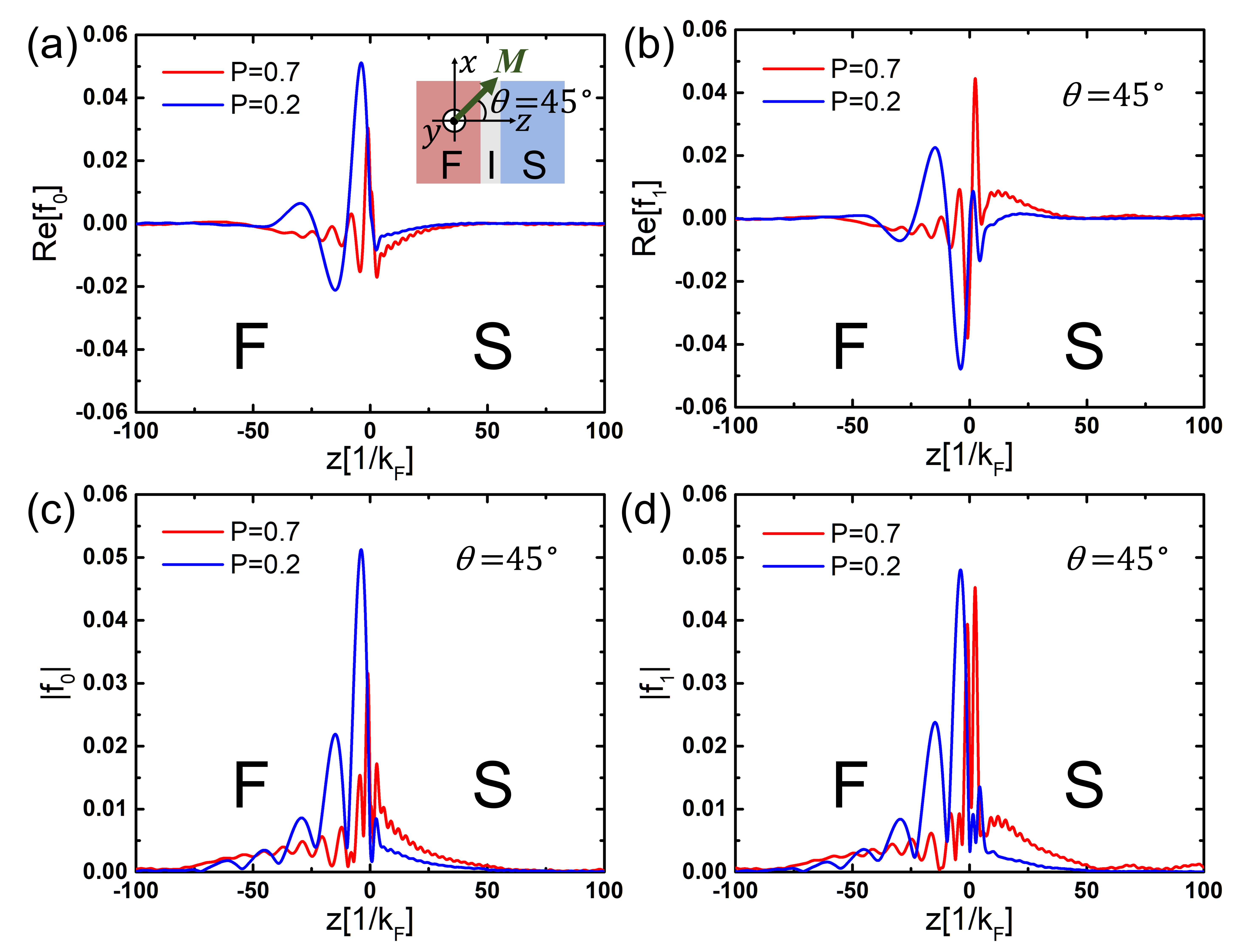}
	\caption{(a), (c) Spatial dependence of the real part and absolute value of the $s_z = 0$ triplet-pair amplitudes for different spin polarizations when $Z=2$, $\lambda=5$ and $\tau = 4$. (b), (d) The same plots for $s_z = \pm 1$ triplet-pair amplitudes.} 
	\label{fig:triplet_P}
\end{figure}

For a weekly spin-polarized F ($P \ll 1$), FFLO correlations in the F region are 
significantly enhanced as compared to when the F region is strongly spin-polarized ($P\sim 1$).
Even in the presence of 
SOC, in Fig.~\ref{fig:triplet_P} we see that such trends are retained on the example of $P=0.2$ and $P=0.7$, respectively.  
Both $f_0$ and $f_1$ for $P=0.2$ are much greater than those for $P=0.7$ in the F region, while this behavior is reversed in the S region. 
For a strong spin polarization, singlet-triplet mixing in the S region become dominant.
The decaying oscillations in the F region have a period  $\propto(k_{\uparrow}-k_{\downarrow})^{-1}$, which is a consequence of the FFLO mechanism~\cite{Demler1997:PRB,Eschrig2015:RPP} and does not require the presence of SOC.
Since $k_{\uparrow}-k_{\downarrow}$ is smaller for $P=0.2$ than for $P=0.7$, in Fig.~9 we see an expected corresponding decrease 
in the period of the decaying oscillations. This also clarifies the changes in the oscillations that we would expect in Figs.~\ref{fig:triplet_in_out}-\ref{fig:triplet_fixing la}, if smaller $P$ values would be considered.

\begin{figure}[t]
	\centering
	\includegraphics*[trim=1cm 0cm 0cm 0cm,clip,width=8.75cm]{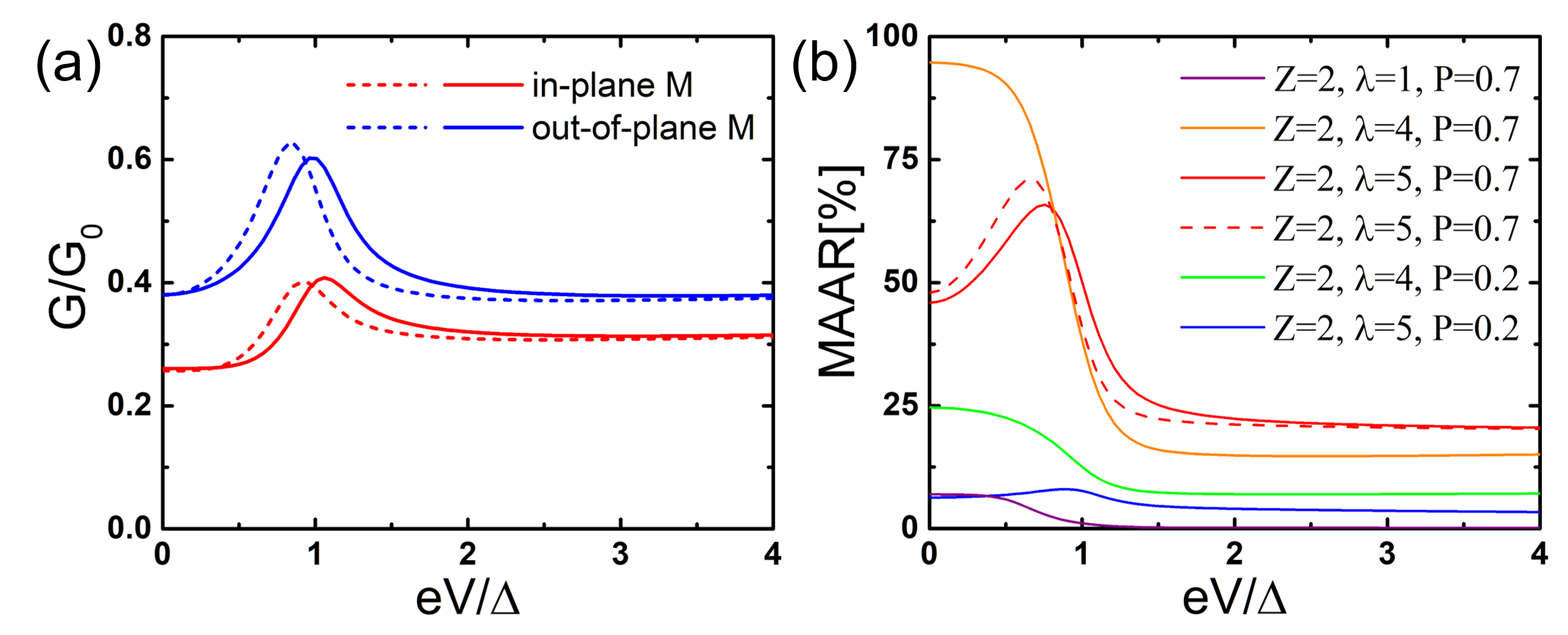}
	\caption{(a) Bias-dependent normalized conductance for $Z = 2$, $\lambda=5$, and $P = 0.7$ with step-function pair potential (solid line) and self-consistent one (dashed line). The blue (red) lines are for an out-of-plane (in-plane) $\bm{M}$. (b) Bias-dependent MAAR for various barrier parameters and spin polarizations. The dashed line indicates the result from the self-consistent pair potential.}
	\label{fig:G_MAAR}
\end{figure}

Following these results for triplet-pair amplitudes, we revisit the study of the interfacial properties on conductance. Unlike considering 
only the step-function $\Delta(z)$ and zero-bias $G$ in Ref.~\cite{Vezin2020:PRB}, in Fig.~\ref{fig:G_MAAR}(a) 
we also show the effect of considering self-consistency and a finite bias. Calculated $G$ for an out-of-plane and in-plane $\bm{M}$ at $T=2K$ shows that the self-consistent $\Delta(z)$ is adequately approximated by the simple step-function pair potential, supporting our
choice in using it to analyze triplet-pair amplitudes, over a large parameter range in $Z$ and $\lambda$.

The most pronounced change is the reduced effective superconducting gap for the self-consistent pair potential, which is a consequence of the decreasing $\Delta(z)$ 
near the interface shown in the Appendix A with Fig.~\ref{fig:self}. While at $T=0\,$K, a step-function approximation leads to $G(V=\Delta)$ (normalized by the Sharvin conductance), 
which only depends on $P$~\cite{Zutic2000:PRB} and not on the interfacial potential barrier, SOC, or the Fermi wave vector mismatch, this universal behavior only
approximately holds for self-consistent $\Delta(z)$, similar as studied in the absence of SOC~\cite{Barsic2009:PRB}.

The conductance favors an out-of-plane  $\bm{M}$ and the resulting MR, as measured in Fig.~\ref{fig:previous_work}, which 
is also referred to as the magnetoanisotropic Andreev reflection (MAAR)~\cite{Hogl2015:PRL,Martinez2020:PRA}
\begin{equation}
	\text{MAAR}(\theta)=[G(0)-G(\theta)]/G(\theta),
	\label{eq:MAAR}
\end{equation}
shown in Fig.~\ref{fig:G_MAAR}(b) is positive. As in Figs.~\ref{fig:scheme} and \ref{fig:previous_work}, the angle $\theta$ is between $\bm{M}$ and the interface normal. 
This MAAR is a superconducting analog of the TAMR~\cite{Fabian2007:APS}, which has the same expression
It is convenient to introduce the amplitude $\text{MAAR}\equiv\text{MAAR}(\theta=\pi/2)$.
The spin-flip Andreev reflection can be considered as a spin-rotation process~\cite{Linder2015:NP}, 
which requires the noncollinearity of the spins between Cooper pairs and the Rashba eigenstates. Since the spins of the Rashba eigenstates lay inside of the interface, an out-of-plane $\bm{M}$ guarantees that no spin-flip Andreev reflection will be suppressed~\cite{Vezin2020:PRB,Cai2021:NC}, 
and thus resulting in a higher conductance. In Fig.~\ref{fig:G_MAAR}(b), we see that the strong spin polarization and suitable barrier parameters lead to a very large MAAR.
The optimal condition for large MAAR is $Z=2, \lambda=4$ or $\lambda = 2Z$, which is very similar to the optimal condition for maximum triplet-pair amplitude. 
The self-consistent calculation reduces the superconducting gap and can slightly increase the subgap MAAR. More importantly, 
MAAR does not always peak at zero bias. It is then possible that experiments which would not focus on zero-bias MAAR~\cite{Cai2021:NC}, 
could reveal an even more enhanced MAAR. In fact, the related results from Fig.~\ref{fig:previous_work} show that the maximum 
MAAR (MR) amplitude already slightly exceeds our simple zero-bias predictions~\cite{Vezin2020:PRB}. 
Even without considering interfacial SOC, the importance of finite-bias conductance results in F/S junctions
has been widely recognized~\cite{Zutic2004:RMP,Zutic1999:PRBa,Zutic2000:PRB,Mazin2001:JAP,Mazin1999:PRL}.

\section{IV. Conclusions and Outlook}
\label{sec:Disc}
With the growing interest in spin-triplet superconductivity, from superconducting spintronics to fault-tolerant quantum computing, there is also a realization that its experimental verification remains a challenge~\cite{Amundsen2024:RMP}. Since many systems suitable for spin-triplet superconductivity also simultaneously support the spin-singlet component, it would be important to understand how such spin-triplet contribution could be enhanced. Our findings for superconducting correlations address both of these issues in F/S junctions. (i) By establishing that various trends in superconducting correlations with interfacial parameters are consistent with trends in the conductance/MR anisotropy, the already existing measurements of such interfacial anisotropy are further justified as a probe of the spin-triplet superconductivity. (ii)  By identifying nonmonotonic trends in the spin-triplet contribution with the interfacial parameters, we offer guidance for a materials design of suitable F/S junctions.
Our approach considers both a simple step-function and a self-consistent form of the pair potential, across a F/S junction.

Superconducting correlations and the zero-bias conductance contribution from equal-spin Andreev reflection represent very different signatures of 
spin-triplet superconductivity. In the first case the spatially-resolved information is obtained from the full-energy range, within the characteristic Debye energy for the  
superconducting pairing, while in the second case the spatial information is absent and only zero-energy behavior is considered. 

Nevertheless, these complementary signatures and the employed different methods point to common nonmonotonic trends in SOC and interfacial barrier for the enhanced spin-triplet superconductivity. Neither vanishingly small interfacial barrier, nor strong SOC, commonly expected to lead to robust proximity effects and strong spin-triplet contribution, are always desirable. Instead, our findings reveal that the dominant influence of spin-triplet superconductivity is realized for the intermediate SOC strength, consistent with the recent experiments on enhanced conductance magnetoanisotropy\cite{Martinez2020:PRA,Cai2021:NC}.

While we have focused on a widely-used Rashba SOC, linear in the wave vector, it would be interesting to explore how other SOC forms, 
cubic in the wave vector~\cite{Alidoust2021:PRB}, or effectively arising from magnetic 
textures~\cite{Gungordu2022:JAP,Kjaergaard2012:PRB,Klinovaja2013:PRL,Kim2015:PRB,Fatin2016:PRL,Yang2016:PRB,Matos-Abiague2017:SSC,Zhou2019:PRB,Desjardins2019:NM,Mohanta2019:PRA,Wei2019:PRL}, 
would modify our findings. 
Would these other SOC forms  
recover enhanced spin-triplet superconductivity which is nonmonotonic in interfacial parameters?  
In two-dimensional systems, there is an also opportunity to consider F/S structures with additional anisotropies and possibly a richer structure
of an inhomogeneous superconductivity~\cite{Samokhin2015:PRB,Diesch2018:NC,Liu2021:S}.
Another direction to explore is realizing tunable SOC, rather than considering multiple samples. 
In planar Josephson junctions this can be implemented using a gate voltage, offering an experimental 
support for the transition between singlet and triplet proximity-induced superconductivity~\cite{Dartiailh2021:PRL,Zhou2022:NC}

For considering more complex structures involving multiple ferromagnetic and superconducting regions, it is very
encouraging that in F/S junctions, as their building block, we already find that SOC-generated spin-triplet superconductivity 
shows long-range effects and that a large magnetic anisotropy of the conductance is experimentally measured 
in heterostructures involving two-dimensional materials~\cite{Cai2021:NC}. A growing number of available 
two-dimensional ferromagnets and semiconductors offers a wealth of opportunities to identify suitable platforms
which support robust spin-triplet superconductivity in their heterostructures. With a large anisotropic MR observed in S/F/S junctions
with all two-dimensional materials~\cite{Kang2021:X}, it would be useful to examine if its origin could be explained by the formation of spin-triplet superconductivity.
Since both the bulk and interfacial SOC is inherent to many superconducting junctions, it would be helpful to revisit some of the prior reports of proximity-induced triplet superconductivity where SOC was not considered~\cite{Sanchez-Manzano2022:NM}, as the missing mechanism for singlet to triplet transformation remains to be identified. Could this transformation be explained by the presence of SOC?

Given the continued interest in identifying spin-triplet superconductivity~\cite{Amundsen2024:RMP}, we expect that the future work will test these trends in interfacial SOC through different 
measurements and examine if they could also guide the design of heterostructures for superconducting spintronics, where the resulting spin currents could be used to control magnetic textures.
Similar design issues pertain to the optimal platforms for topological superconductivity, which usually relies on the equal-spin triplet superconductivity and thus future measuring a large magnetic anisotropy of the conductance may offer a complementary support for topological superconductivity.

\section{Acknowledgements}
We thank Farkhad Aliev, Ranran Cai, and Wei Han for providing their 
experimental data, Chien-Te Wu, \and Oriol T. Valls for valuable discussions on self-consistent calculations. 
This work is supported by U.S. Department of Energy, Office of Science, Basic Energy Sciences 
under Award No. DE- SC0004890. Computational resources were provided by the UB Center for Computational Research. 

\section*{Appendix A: Convergence in Self-Consistent Calculations}
\begin{figure}[h]
	\centering
	\includegraphics*[trim=1.5cm 0.2cm 1.6cm 0.4cm,clip,width=8.8cm]{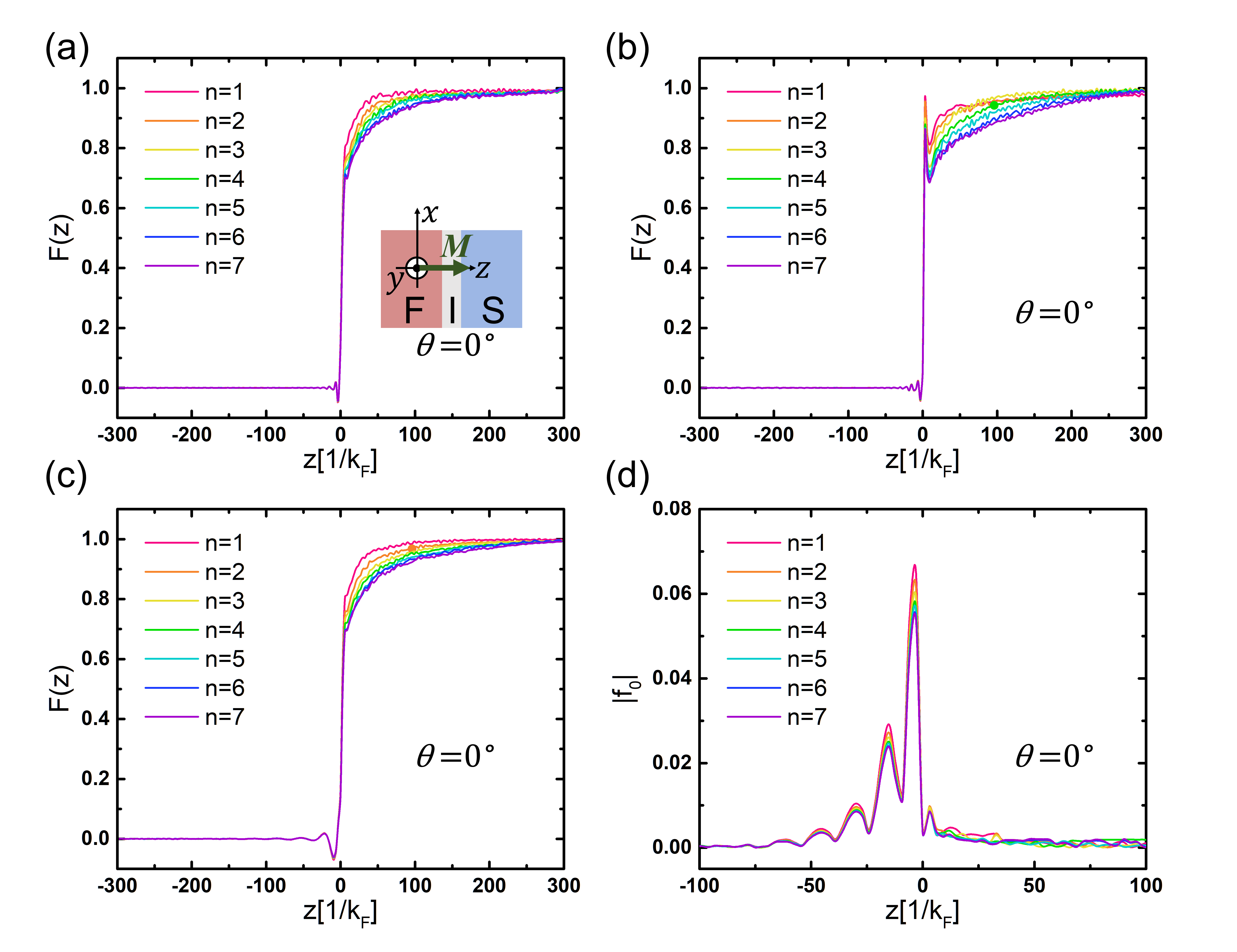}
	\caption{(a) Spatial dependence of the singlet-pair amplitudes for the full self-consistent iterations with $Z=2, \lambda=5, P=0.7$. (b) The same plot with $Z=2, \lambda=10, P=0.7$. (c) The same plot with $Z=2, \lambda=5, P=0.2$. (d) Spatial dependence of the absolute value of the $s_z=0$ triplet-pair amplitudes for the full self-consistent iterations with $Z=2, \lambda=5, \tau=4, P=0.2$. 
	All plots are for an out-of-plane $\bm{M}$.}
	\label{fig:self}
\end{figure}

Most of the self-consistent studies of proximity effects in superconducting junctions employ quasiclassical approximation. The approaches beyond this 
assumption are still relatively rare and computationally more demanding, while their convergence to the self-consistent solution can be a challenge~\cite{Valls:2022}.
In our calculations, using the scattering-state approach and the transfer matrix method, we can achieve a very good convergence of the full self-consistent pair potential 
within 10 iterations.
The corresponding results can be seen in Fig.~\ref{fig:self}. By comparing Fig.~\ref{fig:self}(a) and \ref{fig:self}(c), we can see that the strong spin polarization suppresses the singlet correlation (Cooper pair amplitude, defined in Sec.~II) $F(z)$ in the F region. We have also checked that for $P=0$ we recover the expected results for the N/S junctions~\cite{Valls2010:PRB}). Figure~\ref{fig:self}(b) shows that a peak in $F(z)$ emerges at the F/S interface, when the effective barrier strength is large enough. 
In Fig.~\ref{fig:self}(d), we see that the self-consistent iteration has little impact on the triplet correlation functions. While for $\tau=0$, $f_1$ triplet-correlation vanishes during the self-consistent process, which is a consequence of the Pauli exclusion principle~\cite{Halterman2008:PRB}. Despite these minor changes in the correlation functions 
after the self-consistent iterations, the main 
features and the overall magnitude of the correlation functions are captured by the step-function model. Therefore, we are able to use the step-function model to simplify the computationally demanding analysis for a large range of the interfacial parameters.

\section*{Appendix B: Perturbative Analysis} 
For an in-plane $\bm{M}$, projecting the mixed-spin state $\left| {{\psi _{IP}}} \right\rangle$ in Eq.~\ref{psi_OP} onto the spin-quantization axis along $\bm{z}$, we can obtain
\begin{equation} 
	\begin{gathered}
		\left| {{\psi _{IP}}} \right\rangle  = \left( {\left| {k + Q{ \uparrow _x}, - k + Q{ \downarrow _x}} \right\rangle  - \left| {k - Q{ \downarrow _x}, - k - Q{ \uparrow _x}} \right\rangle } \right) \hfill \\
		= \frac{1}{2}( - \left| {k + Q \uparrow , - k + Q \uparrow } \right\rangle  + \left| {k + Q \uparrow , - k + Q \downarrow } \right\rangle  \hfill \\
		- \left| {k + Q \downarrow , - k + Q \uparrow } \right\rangle  + \left| {k + Q \downarrow , - k + Q \downarrow } \right\rangle  \hfill \\
		+ \left| {k - Q \uparrow , - k - Q \uparrow } \right\rangle  + \left| {k - Q \uparrow , - k - Q \downarrow } \right\rangle  \hfill \\
		- \left| {k - Q \downarrow , - k - Q \uparrow } \right\rangle  - \left| {k - Q \downarrow , - k - Q \downarrow } \right\rangle ). \hfill \\ 
	\end{gathered}
\end{equation} 
We can then calculatee its projection onto the symmetry states and get the results in the main text.

Applying a part of the Rashba SOC as a perturbation to a two-particle
state $\left| {k \sigma , k' \sigma' } \right\rangle$, we obtain
\begin{equation}
	{\hat{k}_y}{{\hat \sigma }_x}\left| {k \sigma , k' \sigma' }
	\right\rangle =  k_y| k \bar\sigma, k' \sigma'\rangle + k'_y| k \sigma ,
	k' \bar\sigma' \rangle,
\label{eq:pert_2nd}
\end{equation}
where $\bar\sigma=-\sigma$ denotes the flipped spin. We then express the first order perturbation for the FFLO states by Rashba SOC, for both out-of-plane and in-plane $\bm{M}$ as proportional to the following state vectors
\begin{widetext}
	\begin{equation}\label{eq:out-pert}
		\begin{gathered}
			\left| {\psi _{OP}^{\left( 1 \right)}} \right\rangle
			\propto  - \alpha {k_y}\left( {\left| {k + Q \uparrow , - k + Q \uparrow } \right\rangle  - \left| {k + Q \downarrow , - k + Q \downarrow } \right\rangle  + \left| {k - Q \uparrow , - k - Q \uparrow } \right\rangle  - \left| {k - Q \downarrow , - k - Q \downarrow } \right\rangle } \right) \hfill \\
			- i\alpha {k_x}\left( {\left| {k + Q \uparrow , - k + Q \uparrow } \right\rangle  + \left| {k + Q \downarrow , - k + Q \downarrow } \right\rangle  + \left| {k - Q \uparrow , - k - Q \uparrow } \right\rangle  + \left| {k - Q \downarrow , - k - Q \downarrow } \right\rangle } \right), \hfill \\ 
		\end{gathered}
	\end{equation}
	
	\begin{equation}\label{eq:in-pert}
		\begin{gathered}
			\left| {\psi _{IP}^{\left( 1 \right)}} \right\rangle
			\propto \alpha {k_y}( - \left| {k + Q \uparrow , - k + Q \uparrow } \right\rangle  + \left| {k + Q \uparrow , - k + Q \downarrow } \right\rangle  - \left| {k + Q \downarrow , - k + Q \uparrow } \right\rangle  + \left| {k + Q \downarrow , - k + Q \downarrow } \right\rangle  \hfill \\
			- \left| {k - Q \uparrow , - k - Q \uparrow } \right\rangle  - \left| {k - Q \uparrow , - k - Q \downarrow } \right\rangle  + \left| {k - Q \downarrow , - k - Q \uparrow } \right\rangle  + \left| {k - Q \downarrow , - k - Q \downarrow } \right\rangle ) \hfill \\
			- i\alpha {k_x}\left( {\left| {k + Q \uparrow , - k + Q \uparrow } \right\rangle  + \left| {k + Q \downarrow , - k + Q \downarrow } \right\rangle  + \left| {k - Q \uparrow , - k - Q \uparrow } \right\rangle  + \left| {k - Q \downarrow , - k - Q \downarrow } \right\rangle } \right). \hfill \\ 
		\end{gathered}
	\end{equation}	
	
\end{widetext}
In considering SOC as a perturbation, we have omitted the term $1/(E_0^{\left( 0 \right)} - {H_0})$, which includes the unperturbed energy and the Hamiltonian, since it only adds coefficients to the perturbed states.
Using these results from the first order perturbation, we are able to give the singlet and triplet pairing components discussed in the main text.

\bibliography{Triplet}
\end{document}